\documentclass[apj]{emulateapj}

\usepackage{graphicx,times}
\usepackage{subfigure}
\newcommand{\be}{\begin{equation}}
\usepackage{threeparttable}
\usepackage{booktabs}
\newcommand{\ee}{\end{equation}}
\newcommand{\bea}{\begin{eqnarray}}
\newcommand{\eea}{\end{eqnarray}}

\usepackage{enumerate}
\usepackage{amsmath}
\usepackage{cases}
\usepackage{longtable}
\usepackage{hyperref}
\usepackage{epstopdf}
\usepackage{amsmath,bm}
\usepackage{amssymb}
\usepackage{natbib}
\usepackage{morefloats}
\usepackage{multirow}
\usepackage{array}
\usepackage{verbatim}

\begin{document}

\title{Nonlinear turbulent dynamo during gravitational collapse}

\author{Siyao Xu\altaffilmark{1,2} and Alex Lazarian\altaffilmark{1,3} }

\altaffiltext{1}{Department of Astronomy, University of Wisconsin, 475 North Charter Street, Madison, WI 53706, USA; 
sxu93@wisc.edu,
lazarian@astro.wisc.edu}
\altaffiltext{2}{Hubble Fellow}
\altaffiltext{3}{Center for Computation Astrophysics, Flatiron Institute, 162 5th Ave, New York, NY 10010}

\begin{abstract}

Via amplification by turbulent dynamo, magnetic fields can be potentially important for the formation of the first stars. 
To examine the dynamo behavior during the gravitational collapse of primordial gas, 
we extend the theory of nonlinear turbulent dynamo to include the effect of gravitational compression.
The relative importance between dynamo and compression varies during contraction, 
with the transition from dynamo- to compression-dominated amplification of magnetic fields with the increase of density.
In the nonlinear stage of magnetic field amplification
with the scale-by-scale energy equipartition between turbulence and magnetic fields, 
reconnection diffusion of magnetic fields in ideal magnetohydrodynamic (MHD) turbulence becomes important. 
It causes the violation of flux-freezing condition and 
accounts for (a) the small growth rate of nonlinear dynamo,
(b) the weak dependence of magnetic energy on density during contraction, 
(c) the saturated magnetic energy,
and (d) the large correlation length of magnetic fields. 
The resulting magnetic field structure and the scaling of magnetic field strength with density 
are radically different from the expectations of flux-freezing.

\end{abstract}

\section{Introduction}

Magnetic fields are ubiquitous in the universe and accompany the cosmic structure formation across cosmic time 
\citep{Beck15,Kron16,Han17,Mar18}. 
They are the important element in many fundamental astrophysical processes, 
but their exact role in star formation remains an open question
\citep{Mckee_Ostriker2007,Cru12, Krum19, Nix19,Pud19,Hen19}.

The formation of the first stars was an epochal event that influenced the subsequent star formation. 
The primordial magnetic fields 
\citep{Bierm50,Qua89,Sigl97,Sch18}, 
if they were amplified via the small-scale turbulent dynamo
\citep{Kaza68,Bran05,XL16},
can affect the primordial star formation and the properties of the first stars 
{\citep{Sur10,SchoSch12,Tur12, Lat13,Mac13,Lat14,Kles19,Shar20}.}

The so-called ``small-scale'' turbulent dynamo with the amplified magnetic fields on scales smaller than the 
driving scale of turbulence has been investigated for decades
\citep{Batc50,Kaza68,KulA92,Subra98,Sch04}.
{The kinematic turbulent dynamo occurs when the magnetic field is weak and its back reaction on the flow is negligible 
\citep{Fede11,Fed16}.
It has an exponential growth of magnetic energy, and the dynamo growth rate depends on plasma parameters, 
including the sonic Mach number
\citep{Fede11}
and the Prandtl number
\citep{Fed14,XL16,Bran19}.}
A long-standing challenge is to formulate the nonlinear turbulent dynamo with significant back-reaction of magnetic fields on turbulence
\citep{Batc50,Sch50,Sub03,SchC07}.
The development of magnetohydrodynamic (MHD) simulations enable numerical studies of nonlinear dynamo 
\citep{Chod00,Hau03,Ryu08,CVB09,Bere11}, 
which reveal (i) a linear-in-time growth of magnetic energy with a universal small growth rate independent of plasma parameters,
as well as (ii) a magnetic energy spectrum following the Kolmogorov spectrum of driven turbulence.

The dynamo theory recently developed by 
\citet{XL16}
(hereafter XL16)
includes multiple physical regimes applicable to a wide range of plasma parameters 
and the nonlinear regime. 
The new kinematic dynamo regime in a weakly ionized plasma predicted by XL16
has been numerically tested with a two-fluid dynamo simulation 
\citep{Xud19}.
The XL16 theory for nonlinear dynamo was established based on recent advances in theoretical 
understanding of strong MHD turbulence 
\citep{GS95,LV99,CV00,MG01,Bere14},
which arises during the nonlinear dynamo. 
As an intrinsic part of MHD turbulence,
the reconnection diffusion of magnetic fields in turbulent media 
dominates over the diffusion of magnetic fields due to microphysical plasma processes
(see \citealt{LV99,Laz05,Sant10,LEC12,Eyin13,Ey15,Kow17,Jaf18,Kow20,San20}, and \citealt{LaR20} for a recent review). 
XL16 for the first time introduced the reconnection diffusion of magnetic fields for nonlinear dynamo. 
Different from earlier models 
\citep{KulA92,Subra98,Sch02}
relying on microscopic diffusion processes, e.g., 
resistive diffusion, ambipolar diffusion,
the XL16 theory for nonlinear dynamo leads to (i) inefficient dynamo growth and (ii) a large correlation length of magnetic fields, 
in agreement with numerical simulations and observations 
\citep{Hau03,Bran05,Ryu08,CVB09,Bere11,XuL17}.

In XL16, we discussed the application of our dynamo formalism to studying the magnetic field amplification during the primordial star formation,
but the effect of gravitational compression was not taken into account. 
Earlier, 
\citet{SchoSch12}
analyzed the turbulent dynamo during the formation of the first stars. 
They included the effect of gravitational compression, as well as non-ideal MHD effects (Ohmic dissipation and ambipolar diffusion), 
but did not consider the reconnection diffusion of magnetic fields. 
A comprehensive study on magnetic fields during the gravitational collapse of primordial gas
has been recently carried out by 
{\citet{Mc20}
(henceforth M+20).}
They provided theoretical predictions on the evolution of magnetic fields and their effects on the first stars, 
and comparisons with predicted numerical outcomes.

For the dynamo process, the initial kinematic stage is usually a transient phase with the magnetic energy smaller than 
the kinetic energy of the smallest turbulent eddies
(see XL16 for the case of a prolonged kinematic stage in a weakly ionized medium),
and it has been carefully investigated in a collapsing primordial gas cloud by M+20.
The subsequent nonlinear stage leads to the energy equipartition between turbulence and magnetic fields
and the increase of correlation length of magnetic fields up to the driving scale of turbulence. 
A proper understanding of the nonlinear dynamo is vital for 
evaluating the influence of magnetic fields on the primordial star formation. 
In this work, on the basis of the XL16 theory
we focus on the nonlinear stage of turbulent dynamo in a gravitationally collapsing system, 
where both dynamo and compression contribute to the growth of magnetic energy. 
We aim to determine the importance of reconnection diffusion for the nonlinear evolution of magnetic fields during gravitational collapse, 
which has not been considered in earlier theoretical works. 
We agree with many aspects discussed in M+20. 
However, our treatment of the diffusion of magnetic fields is different, 
which entails significant differences in the results.
In Section 2, we present the theoretical formulation. 
The comparisons of our results to earlier theoretical and numerical studies 
are discussed in Section 3. 
Our conclusions are in Section 4.

\section{Nonlinear turbulent dynamo during self-gravitational compression}
\label{sec:nonfor}

\subsection{Reconnection diffusion of turbulent magnetic fields}
\label{ssec: recdiff}

Turbulent diffusion of magnetic fields is a part of the classical theory of the 
mean field dynamo  
\citep{Park79}, 
which was invoked to explain the absence of numerous small-scale magnetic field reversals in observations. 
However, 
it only applies to dynamically weak magnetic fields that can be passively advected by hydrodynamic motions.

For the small-scale turbulent dynamo, 
the diffusion of magnetic fields was attributed to Ohmic resistivity in earlier theoretical studies  
(e.g., \citealt{Ruz89,Sch02}). 
However, under this consideration 
the resulting magnetic energy spectrum peaks at the resistive scale, 
which cannot be reconciled with simulations and observations 
\citep{Hau03,Vog05}.

Understanding the diffusivity of magnetic fields in turbulent flows
requires understanding the fundamental process of reconnection of magnetic fields and its dynamical consequence, 
which has been a long-standing puzzle. 
The classical studies of reconnection of non-turbulent magnetic fields were presented in 
\citet{Park57} and \citet{Swe58}, 
but the reconnection rate of the Sweet-Parker model is too slow to explain explosive solar flares
\citep{Par63}.
The model for magnetic reconnection in turbulence was proposed by 
\citet{LV99}
(henceforth, LV99).
Within this model,
the reconnection rate was found to be determined by the turbulent eddy-turnover rate. 
The theory of turbulent reconnection predicts that 
magnetic fields do not constrain turbulent motions that mix them in the direction perpendicular to the local magnetic field. 
Its predictions have been tested
with both simulations and observations
(see \citealt{LaR20}).
According to the LV99 theory, 
turbulent reconnection is a part of MHD turbulent cascade.

For super-Alfv\'{e}nic turbulent motions with the turbulent energy higher than the magnetic energy, 
turbulent motions stretch and amplify magnetic fields with the turbulent energy converted to magnetic energy.
When the magnetic energy reaches equipartition with the turbulent energy, the growing magnetic tension starts to play a dynamically important 
role and suppresses the turbulent stretching. 
Magnetic reconnection acts to release the magnetic tension and convert the magnetic energy to turbulent energy.  
So when the turbulent motions become trans-Alfv\'{e}nic with comparable turbulent and magnetic energies, 
turbulent stretching of magnetic fields is balanced by the reconnection relaxation with shrinking magnetic fields 
on all length scales within the inertial range of turbulence, 
and there is no net growth of either magnetic energy or turbulent energy.

Turbulent reconnection of magnetic fields enables their diffusion (slippage) relative to plasma. 
The corresponding process
termed as ``reconnection diffusion" 
was described in 
\citet{Laz05}
(see also \citealt{Eyink2011, Laz14r,Ey15}), 
and the consequent breakdown of flux-freezing was
numerically demonstrated by 
\citet{Sant10,Eyin13,Lal15}.
For super-Alfv\'{e}nic turbulent motions, the dynamo generation of magnetic fields overwhelms the reconnection diffusion. 
The reconnection diffusion rate $k V_A$, i.e., the rate for shrinking of reconnected magnetic field lines,
is smaller than the dynamo rate (i.e., eddy-turnover rate) $ k v_k$.
For trans-Alfv\'{e}nic turbulent motions, the balance between dynamo generation and reconnection diffusion of magnetic fields is achieved, 
with the reconnection diffusion rate $k_\| V_A$ equal to the dynamo rate $ k_\perp v_k$ 
according to the critical balance relation 
\citep{GS95}.
Here $k_\|$ and $k_\perp$ are the parallel and perpendicular components of wavenumber $k$
\footnote{The caveat here is that $1/k_\|$ and $1/k_\perp$ are the proxies of 
the parallel and perpendicular scales of turbulent eddies, which are 
measured with respect to the local direction of magnetic fields. 
This follows from the eddy description of MHD turbulence in LV99
and was numerically demonstrated by 
\citet{CV00}
and 
\citet{MG01}. 
This notion of local reference system is an important element of the modern theory of MHD turbulence.},
$V_A$ is the Alfv\'{e}n speed, and 
$v_k$ is the turbulent velocity at $k$.
Obviously, for the largest eddy of trans-Alfv\'{e}nic turbulence, 
there is $V_A = v_k$, and thus $k_\| = k_\perp$. 
Reconnection diffusion takes place for both super- and trans-Alfv\'{e}nic turbulent motions. 
In the former case, 
the turbulent dynamics is marginally affected by magnetic fields, and therefore the concept of ``turbulent diffusion" mentioned earlier can be applied. 
The corresponding small-scale kinematic dynamo was formulated by 
\citet{Kaza68}.
For the trans-Alfv\'{e}nic turbulent motions arising in the nonlinear stage of dynamo, 
the back-reaction of magnetic fields and their reconnection diffusion become important, 
which determines the dynamo behavior in the nonlinear regime 
(XL16).

In a gravitationally collapsing system, 
both turbulent dynamo and compression can amplify magnetic fields. 
As long as the amplified magnetic fields reach energy equipartition with turbulence, 
the balance $k_\| V_A = k_\perp v_k$ applies. 
Consequently, magnetic energy cannot grow as expected from the flux-freezing condition because of the loss of magnetic flux via 
reconnection diffusion.

\subsection{Nonlinear dynamo during gravitational collapse}
\label{ssec: genong}

For our analytical model for the gravitational compression, 
we consider an isothermal collapsing sphere with a uniform density distribution
\citep{Spit68}.
The compression factor is defined as 
\begin{equation}
     C = \frac{r}{r_0} ,
\end{equation}
where $r_0$ is the initial radius of the sphere, and $r$ is the radius of the sphere at time $t$. 
Its functional form will be specified later. 
With the mass conservation of the sphere, we find the ratio of the density $\rho$ at $t$ to the initial density $\rho_0$ of the sphere as
\begin{equation}\label{eq: rrhcsp}
     \frac{\rho}{\rho_0}  = \frac{r_0^3}{r^3} = C^{-3}.
\end{equation}
We assume that turbulence is driven in the Jeans-unstable region and 
adopt the Jeans length as the driving scale of turbulence
\citep{Sur10}
\begin{equation}\label{eq: jeanl}
     L = \lambda_J = \sqrt{\frac{\pi }{G\rho}} c_s = L_0 \Big(\frac{\rho}{\rho_0}\Big)^{-\frac{1}{2}}  = L_0 C^\frac{3}{2},
\end{equation}
where $L_0$ is $L$ at $t=0$ and $c_s$ is the sound speed.  
The injected turbulent velocity at $L$ is comparable to $c_s$
\citep{Sur10},
\begin{equation}\label{eq: thevel}
      V_L = c_s. 
\end{equation}
The gravitationally driven turbulence can amplify the seed magnetic field via the turbulent dynamo. 
{The dynamo growth of magnetic energy due to the turbulent shear  
depends on the scaling of turbulent velocities
(see \citealt{Scho12}). }
In the scenario where 
the eddy turnover time of turbulence is much shorter than the contraction time, 
the adiabatic amplification of turbulence with additional enhancement of turbulent velocity due to contraction 
\citep{RobG12,Lee15,Xug20}
can be neglected. 
Here we restrict our analysis to this situation, and thus 
the Kolmogorov scaling of turbulence within the Jeans volume persists during the global collapse.

The initial kinematic stage of dynamo with an exponential growth of magnetic energy has a negligible timescale 
compared with the contraction timescale 
(M+20). 
We note that in the kinematic stage, the reconnection diffusion rate $k V_A$ is smaller than the dynamo rate $k v_k$,
and thus the effect of reconnection diffusion is unimportant. 
At the end of kinematic stage, the magnetic energy is equal to the kinetic energy of the smallest turbulent eddies, 
and the correlation length of magnetic fields is equal to the size of the smallest eddies
(XL16).

The subsequent nonlinear dynamo is characterized by the scale-by-scale equipartition between the magnetic energy $\mathcal{E}$
and the turbulent kinetic energy. 
So there is 
\begin{equation}\label{eq: maeqp}
     \mathcal{E}  = \frac{1}{2} v_p^2 = \frac{1}{2} L^{-\frac{2}{3}} V_L^2 k_p^{-\frac{2}{3}},
\end{equation}
where $v_p$ is the turbulent velocity at $k_p$, 
\begin{equation}\label{eq: kolsc}
    v_p = V_L (k_pL)^{-\frac{1}{3}},
\end{equation}
and $k_p$ is the equipartition wavenumber within the inertial range of turbulence. 
At $k> k_p$, trans-Alfv\'{e}nic MHD turbulence with comparable turbulent and magnetic energies  
\citep{GS95}
is established, 
and the magnetic energy spectrum follows the same Kolmogorov spectrum as turbulence
\citep{Bran05,Bere11}.
The balance between the generation and reconnection diffusion (see Section \ref{ssec: recdiff})
of magnetic fields holds on all length scales smaller than $1/k_p$.
Accordingly, there is no net growth of magnetic energy.

In the super-Alfv\'{e}nic turbulence at $k< k_p$, the turbulent energy is larger than the magnetic energy. 
The dynamo generation dominates over the reconnection diffusion of magnetic fields, resulting in the dynamo growth of magnetic energy.
The turbulent stretching of magnetic fields 
is mainly contributed by the turbulent eddies at $k_p$, with the dynamo stretching rate $\Gamma$ given 
by their eddy turnover rate (Eqs. \eqref{eq: jeanl} and \eqref{eq: kolsc}), 
\begin{equation}\label{eq: dsrate}
   \Gamma = \Gamma_p  = v_p k_p = L^{-\frac{1}{3}} V_L k_p^\frac{2}{3}   
   =   L_0^{-\frac{1}{3}} C^{-\frac{1}{2}} V_L k_p^\frac{2}{3}  ,
\end{equation}
which is higher than that of larger turbulent eddies. 
We note that the eddy turnover rate of turbulence increases with gravitational contraction due to the decrease of length scales. 
By combining Eqs. \eqref{eq: maeqp} and \eqref{eq: dsrate}, we see that 
\begin{equation}\label{eq: Gamme}
    \Gamma \mathcal{E} = \frac{1}{2} L^{-1} V_L^3 =  \frac{1}{2} L_0^{-1} V_L^3  C^{-\frac{3}{2}} = \frac{1}{2} \epsilon_0 C^{-\frac{3}{2}} ,
\end{equation}
where $\epsilon_0 = L_0^{-1} V_L^3$ is the initial energy transfer rate of turbulent energy cascade. 
The scale-independent energy transfer rate 
\begin{equation}\label{eq: tuetrcm}
     \epsilon = L^{-1} V_L^3  = k v_k^3  = \epsilon_0 C^{-\frac{3}{2}} 
\end{equation}
also increases with compression.

The Kazantsev spectrum
\citep{Kaza68,KulA92}
of magnetic energy resulting from the turbulent stretching of magnetic fields takes the form 
\begin{equation}
\begin{aligned}
    M(k,t) &= M_r \exp \Big(\frac{3}{4} \int \Gamma dt \Big)\Big(\frac{k}{k_r}\Big)^\frac{3}{2} \\
              & = M_{r0} C^{\alpha+\frac{3}{2}} \exp \Big(\frac{3}{4} \int \Gamma dt \Big)\Big(\frac{k}{k_{r0} C^{-\frac{3}{2}}}\Big)^\frac{3}{2}  .
\end{aligned}
\end{equation}
{This shape of spectrum has been confirmed by simulations of incompressible/weakly compressible turbulence  
(e.g., \citealt{Mar01,Hau04}), 
as well as simulations of supersonic turbulence
\citep{Fed14}.}
Here the magnetic energy spectrum at the reference wavenumber $k_r$ is 
\begin{equation}
       M_r = \frac{\mathcal{E}_r}{k_r}  = \frac{\mathcal{E}_{r0} C^\alpha}{k_{r0} C^{-\frac{3}{2}}} =  M_{r0} C^{\alpha+\frac{3}{2}},
\end{equation}
with the initial magnetic energy spectrum $M_{r0}$ at the initial reference wavenumber $k_{r0}$.
The reference magnetic energy $\mathcal{E}_r = k_r M_r = \mathcal{E}_{r0} C^\alpha$ increases due to compression,
where $\mathcal{E}_{r0}$ is its initial value. 
If the magnetic flux is perfectly frozen to the gas, 
there is $B\propto \rho^{2/3}$, and thus
the dependence of $\mathcal{E}$ on gas density $\rho$ is 
\begin{equation}\label{eq: scfro}
       \mathcal{E} = \frac{1}{2}  V_A^2 = \frac{B^2}{8\pi\rho} \propto \rho^\frac{1}{3},
\end{equation}
where $B$ is the magnetic field strength. 
Accordingly, the value of $\alpha$ is $-1$ (Eq. \eqref{eq: rrhcsp}) for isotropic compression.
\footnote{{In the case of one-dimensional compression along the magnetic field, $B$ is independent of $\rho$, and $\rho \propto C^{-1}$. 
So there is $\mathcal{E} \propto C$ with $\alpha = 1$. }}

As the dynamo growth of $\mathcal{E}$ happens at $k<k_p$,
$\mathcal{E}$ can also be expressed as the integral of $M(k,t)$ over the wavenumbers smaller than $k_p$, 
\begin{equation}\label{eq: exmks}
\begin{aligned}
    \mathcal{E} &= \frac{1}{2} \int_0^{k_p} M(k,t) dk \\
                       &= \frac{1}{5}\mathcal{E}_{r0} C^{\alpha}  \Big(\frac{k_p}{k_{r0} C^{-\frac{3}{2}}}\Big)^\frac{5}{2} \exp \Big(\frac{3}{4} \int \Gamma dt \Big).
\end{aligned}
\end{equation}
Given the expressions of $\mathcal{E} $ from both Eqs. \eqref{eq: maeqp} and \eqref{eq: exmks}, 
now we are able to obtain the time evolution of $\mathcal{E} $.
By applying $d \ln / dt$ to both sides of Eqs. \eqref{eq: maeqp} and \eqref{eq: exmks}, we get
\begin{equation}
   \frac{d \ln \mathcal{E} }{dt} = -\frac{d \ln C}{dt} - \frac{2}{3} \frac{d \ln k_p}{dt},
\end{equation}
and
\begin{equation}
    \frac{d \ln \mathcal{E}}{ dt} = (\alpha + \frac{15}{4}) \frac{d \ln C}{dt} + \frac{5}{2} \frac{d \ln k_p}{dt}  +\frac{3}{4} \Gamma,   
\end{equation}
respectively. 
By combining the above two expressions and using Eq. \eqref{eq: Gamme}, we find 
\begin{equation}
    \frac{d \mathcal{E}}{dt} = \frac{4\alpha}{19} \mathcal{E} \frac{d \ln C}{ dt} + \frac{3}{38}  \epsilon_0 C^{-\frac{3}{2}} .
\end{equation}
It has the solution 
\begin{equation}\label{eq: gelemen}
       \mathcal{E} =  \frac{3}{38}  \epsilon_0 C^\frac{4\alpha}{19} \int_{t_\text{cr}}^t C^{-\frac{3}{2} - \frac{4\alpha}{19}} dt  + \mathcal{E}_\text{cr} C^\frac{4\alpha}{19},
\end{equation}
where $\mathcal{E}_\text{cr}$ is the magnetic energy at the onset of nonlinear dynamo at $t = t_\text{cr}$. 
At $C=1$, it recovers the formula of nonlinear dynamo without compression in XL16, 
\begin{equation}\label{eq: xl16nc}
      \mathcal{E} = \frac{3}{38} \epsilon_0 (t-t_\text{cr}) + \mathcal{E}_\text{cr},
\end{equation}
with the dynamo growth rate $3/38 \epsilon_0$ more than an order of magnitude smaller than the 
turbulent energy transfer rate. 
The factor $3/38$ shows the low efficiency of nonlinear dynamo due to the reconnection diffusion of magnetic fields
\footnote{
Incidentally 
\citet{KulA92}
also derived $3/38 \epsilon_0$ as the rate of dynamo 
by assuming that ``the power driving both the reconnection and growth of magnetic noise becomes comparable to the turbulent power".
However, this assumption cannot be physically justified with the Petschek's model for reconnection 
\citep{Pet64}
adopted there.}, 
which is consistent with the results of numerical simulations 
\citep{CVB09,Bere11}.
The second term on the RHS of Eq. \eqref{eq: gelemen} describes the growth of magnetic energy solely due to compression. 
It shows that because of the reconnection diffusion of magnetic fields and thus the violation of flux-freezing in the nonlinear regime, 
the growth of magnetic energy due to compression is also inefficient with $\mathcal{E} \propto C^\frac{4\alpha}{19}$
instead of $\mathcal{E} \propto C^\alpha $.

Besides the time evolution of $\mathcal{E}$,
given the relation between $\mathcal{E}$ and $k_p$ in Eq. \eqref{eq: maeqp}, 
we can also obtain $k_p$ as a function of $t$ during the nonlinear dynamo by using the result in Eq. \eqref{eq: gelemen}, 
\begin{equation}
     k_p = \Bigg[\frac{3}{19}\epsilon_0^\frac{1}{3} C^{1+\frac{4 \alpha}{19}} \int_{t_\text{cr}}^t C^{-\frac{3}{2} - \frac{4\alpha}{19}} dt  + 2 \epsilon_0^{-\frac{2}{3}} \mathcal{E}_\text{cr} C^{1 + \frac{4\alpha}{19}}  \Bigg]^{-\frac{3}{2}}.
\end{equation}
$k_p^{-1}$ is the correlation length of the amplified magnetic field, 
which becomes comparable to $L$ at the full saturation of nonlinear dynamo.

\subsection{Nonlinear dynamo during free-fall collapse}
\label{ssec: ndff}

To further examine the evolution of $\mathcal{E}$ under the effects of both dynamo and compression, 
we follow M+20 and adopt the model for the 
free-fall collapse of a uniform sphere that is initially at rest 
\citep{Spit68}.
The compression factor in this scenario can be approximated by 
(M+20; see also \citealt{Giri14}),
\begin{equation}\label{eq: spitzc}
     C = \frac{r}{r_0} \approx \Bigg(1- \Big(\frac{t}{t_{ff}}\Big)^2\Bigg)^\frac{2}{3} ,
\end{equation}
where 
\begin{equation}\label{eq: tfref}
    t_{ff} = \sqrt{\frac{3\pi}{32 G \rho_0}}
\end{equation}
is the initial free-fall time of the sphere, and $G$ is the gravitational constant. 
The free-fall collapse starts very slowly, and then the contraction becomes ever faster. 
Although this model does not include the pressure support, 
it well describes the initial stage of gravitational collapse especially when the mass exceeds the Jeans mass
\citep{Vog16}.

Given the expression of $C$, we see that the integral in Eq. \eqref{eq: gelemen} becomes 
\begin{equation}\label{eq: intneum}
\begin{aligned}
     \int_{t_\text{cr}}^t C^{-\frac{3}{2} - \frac{4\alpha}{19}} dt 
   &  \approx  \int_{t_\text{cr}}^t \Bigg(1- \Big(\frac{t}{t_{ff}}\Big)^2\Bigg)^{-  \frac{2}{3}   (\frac{3}{2} + \frac{4\alpha}{19} )} dt \\
   & = t_{ff} \int_{\frac{t_\text{cr}}{t_{ff}}}^{\frac{t}{t_{ff}}} \Bigg(1- \Big(\frac{t}{t_{ff}}\Big)^2\Bigg)^\beta d \Big(\frac{t}{t_{ff}}\Big) ,
\end{aligned}
\end{equation}
where 
\begin{equation}
    \beta = -  \frac{2}{3}   \Big(\frac{3}{2} + \frac{4\alpha}{19} \Big).
\end{equation}
At a short time, i.e., $t \ll t_{ff}$, the contraction is insignificant with $C \approx 1$. 
Hence the above integral has the asymptotic form as
\begin{equation}
     \int_{t_\text{cr}}^t C^{-\frac{3}{2} - \frac{4\alpha}{19}} dt 
    \approx t - t_\text{cr}.
\end{equation}
Then the formula given by Eq. \eqref{eq: xl16nc} applies. 
It means that at the initial stage of collapse, the growth of magnetic energy mainly comes from turbulent dynamo.

We further write the expression of $\mathcal{E}$, normalized by the turbulent energy $V_L^2/2$ at $L$,
(Eqs. \eqref{eq: jeanl}, \eqref{eq: thevel}, \eqref{eq: tuetrcm}, \eqref{eq: gelemen}, \eqref{eq: tfref}, and \eqref{eq: intneum}),
\begin{align}
     \frac{\mathcal{E}}{ \frac{1}{2} V_L^2} &\approx \frac{3}{19} \sqrt{\frac{3}{32}} C^\frac{4\alpha}{19}  \int_{\frac{t_\text{cr}}{t_{ff}}}^{\frac{t}{t_{ff}}} \Bigg(1- \Big(\frac{t}{t_{ff}}\Big)^2\Bigg)^\beta d \Big(\frac{t}{t_{ff}}\Big)    \nonumber\\
     & ~~~~ + \frac{\mathcal{E}_\text{cr}}{\frac{1}{2} V_L^2} C^\frac{4\alpha}{19}   \label{eq: genumrat}\\
     & \approx \frac{3}{19} \sqrt{\frac{3}{32}} C^\frac{4\alpha}{19} \sqrt{1-C^\frac{3}{2}} + \frac{\mathcal{E}_\text{cr}}{\frac{1}{2} V_L^2} C^\frac{4\alpha}{19}  \label{eq: ratsht} .
\end{align}
Eq. \eqref{eq: ratsht} is its approximate form at a short time,
where we use the relation 
\begin{equation}
     t = t_{ff} \sqrt{1-C^\frac{3}{2}}
\end{equation}
derived from Eq. \eqref{eq: spitzc}, 
and we assume $t_\text{cr} / t_{ff} \approx 0 $ given the negligible timescale of kinematic dynamo compared to $t_{ff}$
(M+20).

\begin{figure*}[htbp]
\centering   
\subfigure[]{
   \includegraphics[width=8.5cm]{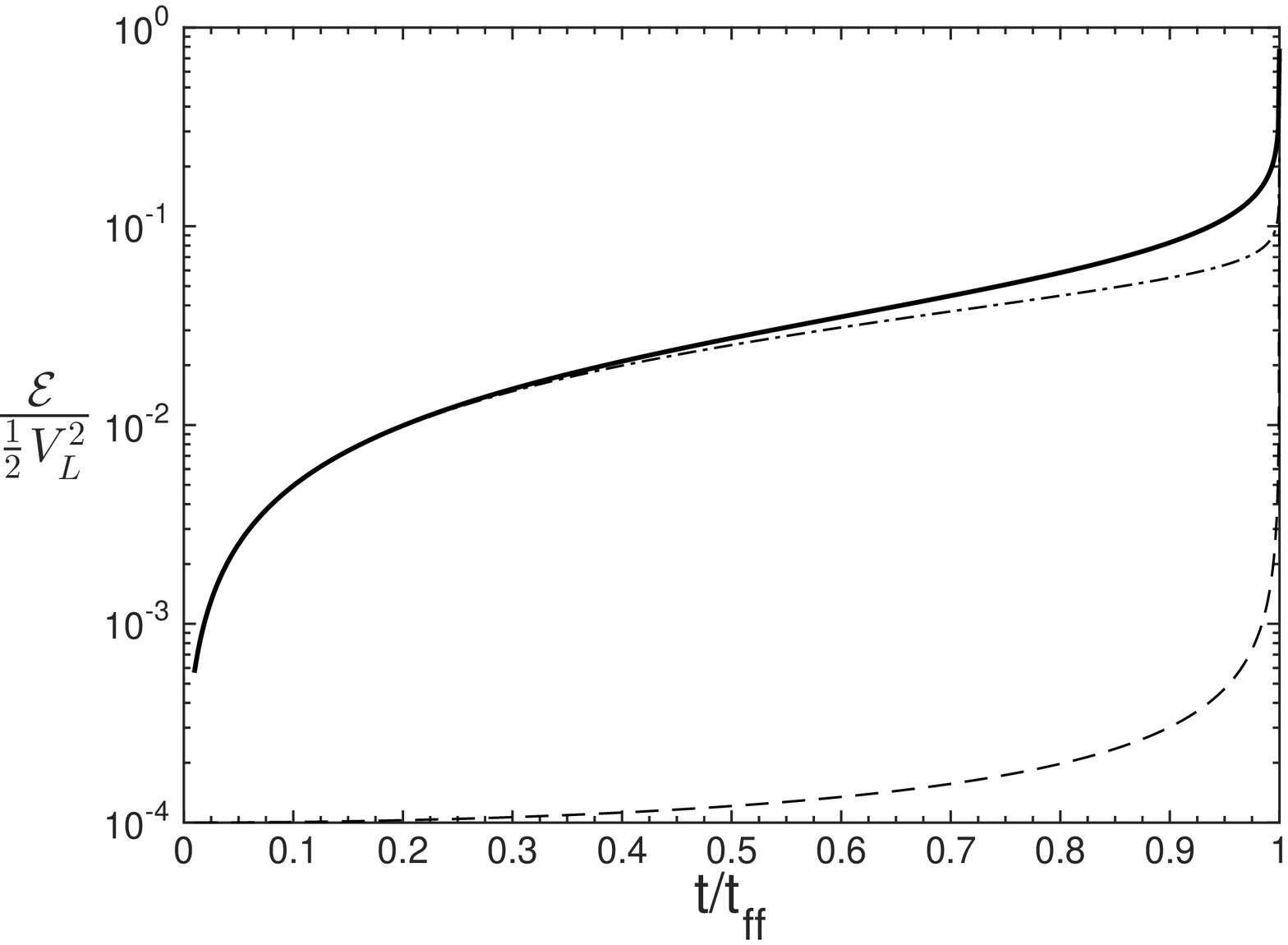}\label{fig: enetim}}
\subfigure[]{
   \includegraphics[width=8.5cm]{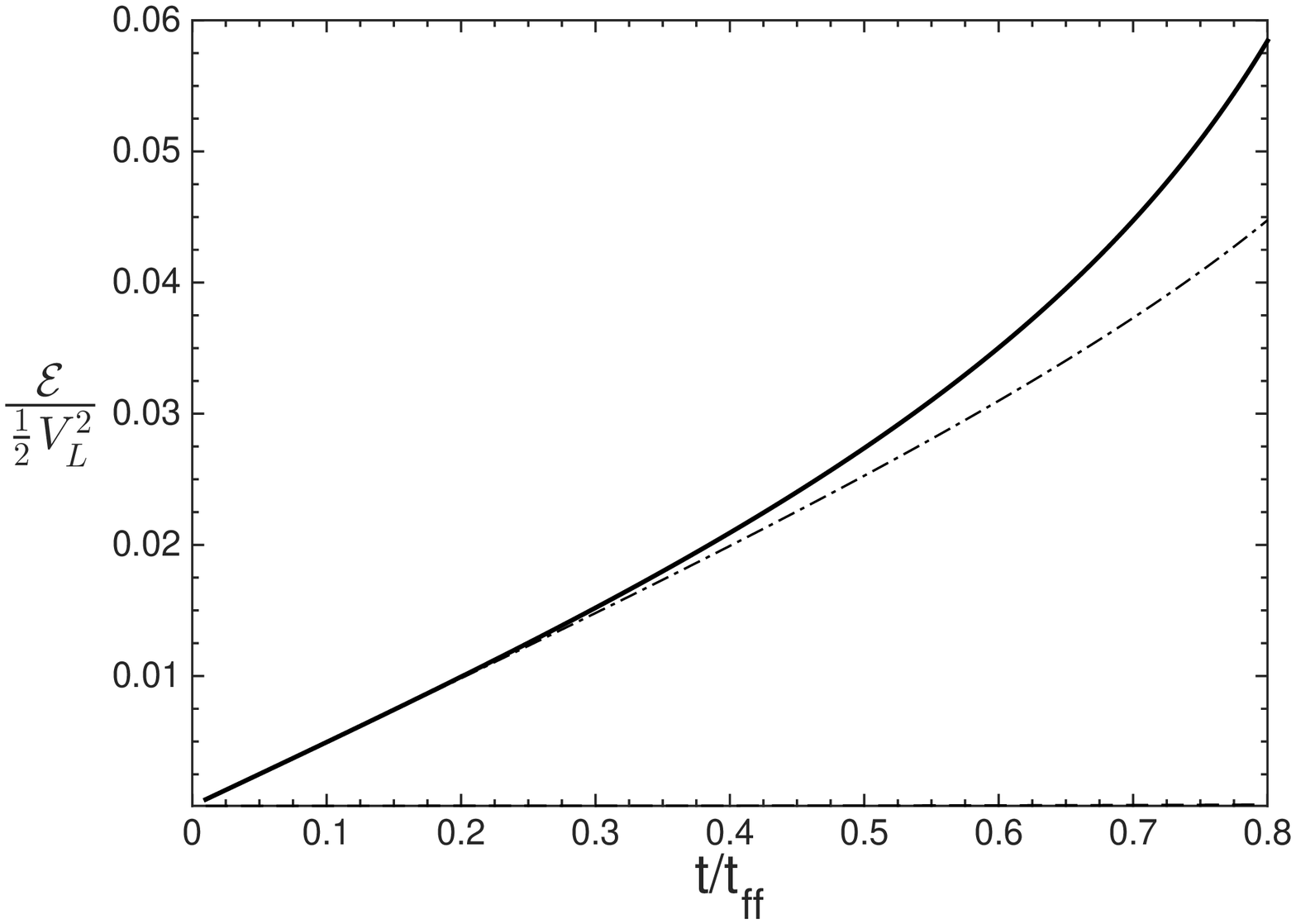}\label{fig: enetimli}}
\caption{ (a) Normalized $\mathcal{E}$ as a function of normalized $t$. The solid line corresponds to 
Eq. \eqref{eq: genumrat}. The dash-dotted line corresponds to its approximation given by 
Eq. \eqref{eq: ratsht}. 
{For comparison}, the dashed line indicates the scaling under the freezing-in condition
(Eq. \eqref{eq: refffs}).
(b) Zoom of (a) for a shorter range of $t$.} 
\label{fig: ene}
\end{figure*}

In Fig. \ref{fig: ene}, we present the normalized $\mathcal{E}$ 
as a function of $t/t_{ff}$.
In this illustration, we adopt ${\mathcal{E}_\text{cr}}/({{1}/{2} V_L^2}) = 10^{-4}$
at the onset of nonlinear dynamo, 
which has a negligible value and does not affect the dynamo behavior. 
From the zoom-in in Fig. \ref{fig: enetimli}, we indeed see the initial linear-in-time growth of $\mathcal{E}$ as dictated by the nonlinear dynamo. 
According to Eq. \eqref{eq: xl16nc},
when there is no gravitational compression, 
it takes $19/3 \approx 6$ largest eddy-turnover time for the nonlinear dynamo to reach the final 
energy equipartition, i.e., 
${\mathcal{E}}/({{1}/{2} V_L^2}) = 1$. 
However, in a collapsing sphere with limited free-fall time, 
the full equipartition with turbulence 
cannot be reached via the nonlinear turbulent dynamo alone.

\begin{figure}[htbp]
\centering   
   \includegraphics[width=8.5cm]{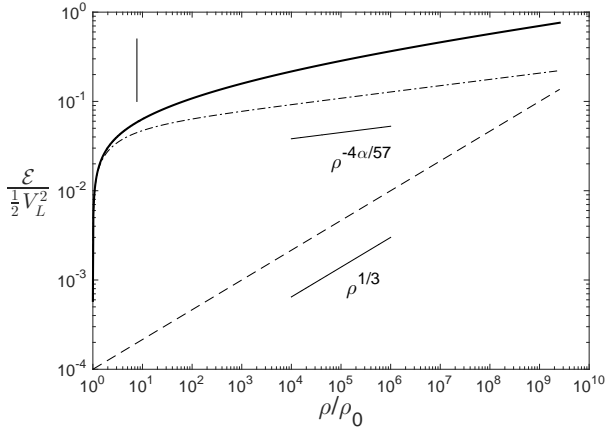}
\caption{Same as Fig. \ref{fig: enetim} but for the normalized $\mathcal{E}$ as a function of normalized $\rho$. 
The vertical line corresponds to $t/t_{ff} = 0.8$. }
\label{fig: enerat}
\end{figure}

The deviation from the dynamo growth takes place at a later time due to the effect of compression. 
In Fig. \ref{fig: enerat}, we present the normalized $\mathcal{E}$ 
as a function of $\rho/\rho_0$ by using the relation in Eq. \eqref{eq: rrhcsp}.
The vertical line indicates the density value 
corresponding to $t/t_{ff} = 0.8$.
When the change in $\rho$ is small, 
the growth of $\mathcal{E}$ is dominated by the nonlinear dynamo, and
$\mathcal{E}$ increases sharply with $\rho$.
As $\rho$ increases rapidly toward the end of collapse,
the growth of $\mathcal{E}$ mainly results from the compression in the later stage of collapse, with the scaling 
slightly steeper than 
\begin{equation}\label{eq: deerdw}
     \mathcal{E} \propto C^\frac{4\alpha}{19} \propto \rho^{-\frac{4\alpha}{57}}.
\end{equation}
The weak dependence of $\mathcal{E}$ on $\rho$ originates from the 
reconnection diffusion of magnetic fields, 
which causes the  leakage of magnetic flux during compression
\citep{Sant10}.
It is important to stress that as shown in Section \ref{ssec: genong}
(see Eq. \eqref{eq: gelemen}), 
the value $4\alpha/19$ is derived by using the Kolmogorov scaling of turbulence 
and the slope of Kazantsev magnetic energy spectrum 
(see e.g., \citealt{Krai67,Eyi10}
for a different slope of Kazantsev spectrum).
It does not depend on the detailed model for collapse and can be generally applied to different scenarios of collapse 
apart from the free-fall collapse considered here.

As a comparison, we also show the result (dashed line) expected from compression alone under the freezing-in condition, i.e., 
\begin{equation}\label{eq: refffs}
    \frac{\mathcal{E}}{ \frac{1}{2} V_L^2}  = \frac{\mathcal{E}_\text{cr}}{ \frac{1}{2} V_L^2} C^\alpha.
\end{equation}
The corresponding growth of $\mathcal{E}$ is insignificant at an early time due to the initially slow contraction (see Fig. \ref{fig: ene}), 
but $\mathcal{E}$ changes steeply with $\rho$ following the scaling in Eq. \eqref{eq: scfro}.

As the actual growth of $\mathcal{E}$ with $\rho$ is very slow, 
to reach the final saturation with $\frac{\mathcal{E}}{\frac{1}{2} V_L^2} = 1$,
an increase in density by many orders of magnitude is required (see Fig. \ref{fig: enerat}). 
When the final saturation is approached, 
the balance between the generation of magnetic fields via both dynamo and compression 
and the reconnection diffusion of magnetic fields exists on all length scales within the inertial range of turbulence. 
As a result, there is no further net growth of $\mathcal{E}$.

\begin{figure}[htbp]
\centering   
   \includegraphics[width=8.5cm]{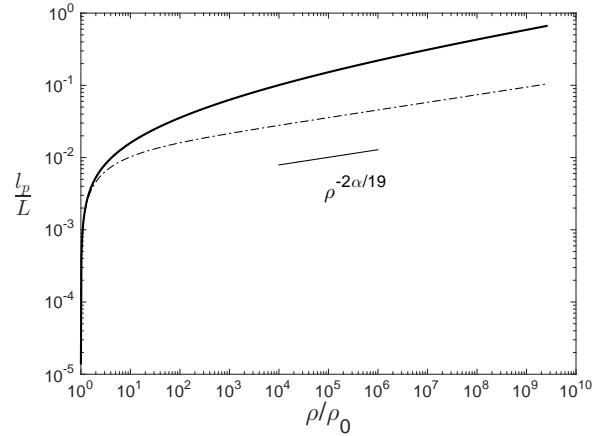}
\caption{ Normalized $l_p$ as a function of normalized $\rho$. }
\label{fig: enelp}
\end{figure}

Given the functional form of $\mathcal{E}$, 
we find the correlation length $l_p = 1/ k_p$ of magnetic field normalized by $L$ as (Eq. \eqref{eq: maeqp})
\begin{equation}
 \frac{l_p}{L} = \Big(\frac{\mathcal{E}}{\frac{1}{2}V_L^2}\Big)^\frac{3}{2}.
\end{equation}
As shown in Fig. \ref{fig: enelp},
following the growth of $\mathcal{E}$, $l_p/L$ first increases sharply with $\rho$ 
and then gradually increases with the scaling slightly steeper than (Eq. \eqref{eq: deerdw})
\begin{equation}
     \frac{l_p}{L} \propto \mathcal{E}^\frac{3}{2} \propto  \rho^{-\frac{2\alpha}{19}}.
\end{equation}
In the saturated state, 
$l_p$ remains comparable to the outer scale of turbulence.

In brief, during the free-fall collapse, 
in the nonlinear stage with energy equipartition between turbulence and magnetic fields, 
the growth of $\mathcal{E}$ is initially attributed to the dynamo amplification 
and then dominated by compression toward the final stage of collapse. 
The efficiency of nonlinear dynamo and the scaling of $\mathcal{E}$ with density depends on the reconnection diffusion of magnetic fields. 
The growth ceases when $\mathcal{E}$ becomes comparable to the kinetic energy of the largest eddy 
and $l_p$ reaches $L$.

\section{Discussion}

\subsection{Comparison with the limit of flux-freezing}

In the kinematic regime with the magnetic energy lower than the turbulent energy of the smallest eddies, 
reconnection diffusion of magnetic fields is insignificant, 
and the approximation of flux-freezing can hold on scales larger than the dissipation scale $1/k_d$ of magnetic fluctuations. 
It follows that the Kazantsev spectrum of magnetic energy can be preserved during compression.
With the spectrum peaked at $k_d$, the magnetic energy is concentrated at small scales. 
The magnetic fields are organized in a folded structure with the length comparable to the flow scale and field reversals at $1/k_d$ 
\citep{Subra98,Schek02,Sch04}. 
The magnetic folds are intermittent in space with a small volume filling fraction. 
The magnetic field amplification due to compression alone has the scaling $B \propto \rho^{2/3}$.

In the nonlinear regime, 
flux freezing breaks down due to reconnection diffusion of magnetic fields. 
The fast turbulent reconnection with the reconnection rate equal to the turbulent eddy-turnover rate does not allow for the 
folded structure of magnetic fields. 
As derived in Section \ref{sec:nonfor},
in the nonlinear stage, given the relations in Eq. \eqref{eq: deerdw} and $\mathcal{E} \propto B^2/\rho$,
there is $B \propto \rho^{2/57+1/2}$ at $\alpha=-1$ purely due to compression. 
After saturation, $\mathcal{E}$ remains constant if the turbulent energy does not change, 
and thus $B$ varies as $B \propto \rho^{1/2}$.

The differences in magnetic field properties between the cases with reconnection diffusion and flux-freezing are summarized in Table \ref{tab:com}.
Clearly, reconnection diffusion plays a key role
in shaping the magnetic field structure and regulating the growth of magnetic fields.  
It leads to a new paradigm for the magnetic field amplification during the primordial star formation,
which radically differs from the expectations in the flux-freezing limit.

\begin{table*}[!htbp]
\renewcommand\arraystretch{1.5}
\centering
\begin{threeparttable}
\caption[]{Comparison between the cases with reconnection diffusion and flux-freezing}\label{tab:com} 
  \begin{tabular}{c|c|c}
     \toprule
                         &            Reconnection diffusion                    &  Flux-freezing  \\
                      \hline
     \multirow{2}{*}{Magnetic energy spectrum}        &         Kazantsev spectrum ($k<k_p$)       &  \multirow{2}{*}{Kazantsev spectrum ($k<k_d$)}   \\
                                                                                &          Kolmogorov spectrum ($k_p<k<k_d $)   &   \\                                   
                      \hline
     Magnetic field structure                       &    Turbulent structure               & Folded structure   \\
                      \hline
     Correlation length of magnetic fields                              &     $1/k_p$ (equipartition scale)      &   $1/k_d$ (dissipation scale)   \\              
                      \hline
      \multirow{2}{*}{Dependence of $B$ on $\rho$ under compression}       &  $B \propto \rho^{\frac{2}{57}+\frac{1}{2}}$ (nonlinear)     &  \multirow{2}{*}{$B\propto \rho^{2/3}$}     \\
                                                                                                          &  $B\propto \rho^\frac{1}{2}$ (saturated)  &    \\
     \bottomrule
    \end{tabular}
 \end{threeparttable}
\end{table*}

\subsection{Comparison with previous analytical work}

In earlier analytical studies of magnetic field amplification during gravitational collapse, 
reconnection diffusion was not taken into account. 
The nonlinear dynamo without reconnection diffusion has the dynamo growth rate comparable to the turbulent 
energy transfer rate 
\citep{Sch02}.
By contrast, we recall that the dynamo growth rate with reconnection diffusion is more than an order of magnitude smaller than the turbulent energy transfer rate.
The nonlinear dynamo without reconnection diffusion
was applied to the context of primordial star formation by
\citet{SchoSch12}, 
where they also included the effect of gravitational compression under the consideration of flux freezing. 
As a result, both the dynamo and compressional amplification of magnetic fields in \citet{SchoSch12}
are much more efficient compared to our results.

The nonlinear effect of Lorentz force on diffusion of magnetic fields was discussed in 
\citet{Sub99}. 
\citet{Sub99} introduced an effective magnetic diffusivity, which depends on the magnetic energy density and can dominate over the 
resistivity when the magnetic field becomes sufficiently strong. 
This model does not involve magnetic reconnection in turbulence, 
and the resulting diffusion of magnetic fields is different from reconnection diffusion. 
Accordingly, its application to nonlinear dynamo 
\citep{Scho15}
leads to different dynamo behavior and magnetic energy spectrum from those in XL16.

M+20 comprehensively analyzed the magnetic fields in the formation of the first stars, 
including their generation via the Biermann battery process, 
and their amplification via kinematic turbulent dynamo, nonlinear turbulent dynamo, and gravitational compression. 
They adopted the XL16 theory for describing the nonlinear dynamo, 
but calculated the compressional amplification of magnetic fields under the flux-freezing condition. 
Therefore, the scaling of $B$ with $\rho$ that they derived is different from ours. 
M+20 pointed out that after reaching full saturation, the magnetic energy remains in equipartition with the turbulent energy
and does not grow further with compression. 
We agree on this statement and identify its physical origin as reconnection diffusion. 
We argue that this energy equipartition is maintained by the balance between the dynamo generation and reconnection 
diffusion of magnetic fields.

\subsection{Comparison with simulations }
\label{ssec:disim}

As a common problem of numerical studies on magnetic field amplification during the primordial star formation, 
it is impossible to 
fully resolve the turbulent cascade that spans many orders of magnitude in length scales with current simulations. 
As the dynamo rate is determined by the eddy-turnover rate, which increases toward smaller scales, 
it is found that 
a minimum resolution between 32 zones per Jeans length 
\citep{Feder11}
and 64 zones per Jeans length 
\citep{Tur12}
is required to properly resolve turbulent motions and capture dynamo action. 
Despite the success in showing the presence of dynamo,
the numerical treatment of dynamo in a gravitationally collapsing system is still problematic. 
As the smallest turbulent eddies at the realistic viscous scale are usually underresolved, 
the kinematic stage of dynamo is unrealistically prolonged, and thus 
it takes a large fraction of the collapse timescale to reach the nonlinear stage
(\citealt{SchoSch12}; M+20). 
Since the effect of compression is already significant at the onset of nonlinear dynamo, 
it dominates over the dynamo effect in amplifying magnetic fields during the entire nonlinear stage. 
Therefore, the nonlinear dynamo behavior in the initial stage of slow collapse as shown in Fig. \ref{fig: enetimli}
is not expected in simulations.

In addition, 
the reconnection diffusion of magnetic fields only becomes important in the nonlinear stage with energy equipartition between 
turbulence and magnetic fields. 
In the numerical cases with unresolved turbulence or the magnetic energy smaller than the turbulent energy of the smallest eddies, 
the reconnection diffusion has a minor effect on growth of magnetic energy, 
which can follow the scaling expected from flux-freezing.

\begin{figure*}[htbp]
\centering   
   \includegraphics[width=14cm]{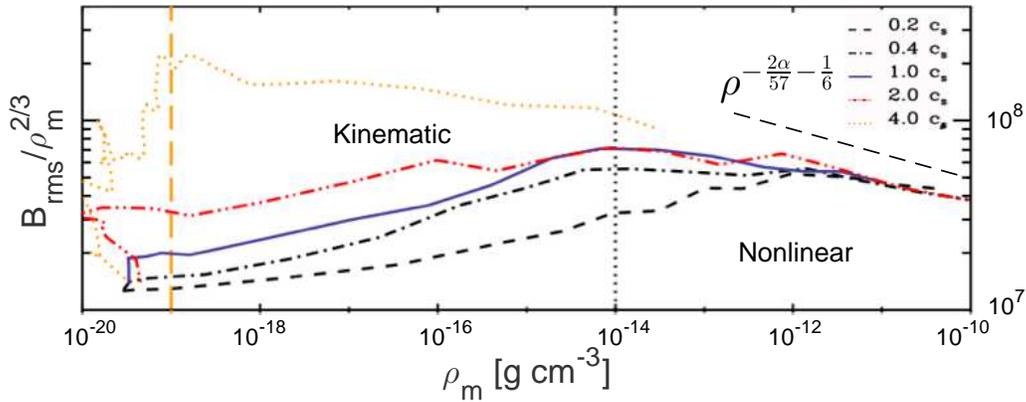}
\caption{ Evolution of $B_\text{rms} / \rho_m^{2/3} $ as a function of $\rho_m$. The original plot is taken from 
\citet{Sur12}. 
The added dashed line in the nonlinear stage corresponds to our analytical prediction given by Eq. \eqref{eq: oupred}. }
\label{fig: fedmod}
\end{figure*}

To compare our theory of nonlinear dynamo in a gravitationally collapsing system with simulations, 
here we take the numerical experiment with ideal MHD simulations carried out by 
\citet{Sur12}
as an example, 
where as shown by the time evolution of magnetic energy spectrum, 
the nonlinear stage is reached. 
Fig. \ref{fig: fedmod} shows the evolution of the rms magnetic field strength 
as a function of the mean density in the central Jeans volume for runs with different initial turbulent velocities in 
\citet{Sur12}. 
These simulations resolve the Jeans length with 64 cells. 
The dotted vertical line in their plot indicates the transition from kinematic to nonlinear stage of magnetic energy growth. 
As expected for simulations, 
the kinematic stage takes a significant fraction of the collapse time, and in the subsequent nonlinear stage, the magnetic field 
amplification is dominated by gravitational compression. 
In the nonlinear stage with significant compression effect, 
our theoretical finding in Section \ref{ssec: ndff} shows that the scaling of $B$ with density is approximately (Eq. \eqref{eq: deerdw})
\begin{equation}
      B \propto \rho^{-\frac{2\alpha}{57}+\frac{1}{2}}.
\end{equation}
When normalized by $\rho^{2/3}$, it is 
\begin{equation}\label{eq: oupred}
      \frac{B}{\rho^\frac{2}{3}} \propto \rho^{-\frac{2\alpha}{57}-\frac{1}{6}}.
\end{equation}
The above scaling with $\alpha = -1$ is indicated by the added dashed line in Fig. \ref{fig: fedmod}, 
which agrees well with their numerical result, 
suggestive of the importance of reconnection diffusion and the breakdown of flux-freezing.

\section{Conclusions}

Based on our earlier study of nonlinear turbulent dynamo, 
here we focus on the nonlinear stage of magnetic field amplification in a gravitationally collapsing system. 
Our main findings include:

1. The nonlinear dynamo, despite its small growth rate, 
can still dominate the magnetic field amplification when the contraction is slow. 
When the contraction is fast, compressional amplification is manifested and dominates over the dynamo amplification.

2. Because of the reconnection diffusion of magnetic fields at the energy equipartition between magnetic fields and turbulence, 
the growth of magnetic energy due to both nonlinear dynamo and gravitational compression are much less efficient 
compared to the case with flux-freezing. 
As a result, the timescale of nonlinear dynamo is much longer than that of free-fall collapse. 
Moreover, to reach the final equipartition between magnetic energy and the turbulent energy of the largest eddy via compression, 
an increase in density by many orders of magnitude is needed.

3. Under the effects of both dynamo and compression, 
the maximum magnetic energy is limited by the turbulent energy of the largest eddy. 
The largest correlation length of magnetic fields is determined by the size of the largest eddy.

The above findings are important for studying the impact of magnetic fields on the first star formation. 
The simulations by 
\citet{Shar20}
suggest that 
the magnetic fields amplified during the collapse of primordial clouds can potentially influence the 
initial mass function of the first stars, 
as strong magnetic fields suppress fragmentation and reduce the number of low-mass stars.
Given the differences between theories and simulations arising from the limited numerical resolution (see Section \ref{ssec:disim}),
the detailed physics induced by highly-resolved turbulence and the  
consequences for the properties of the first stars should be further investigated.

In this work we consider the Kolmogorov turbulence in a relatively slow contraction with homogeneous density distribution. 
In a more realistic collapsing environment, 
when the local contraction time of the central high-density region 
becomes shorter than the eddy turnover time, 
turbulent eddies are adiabatically compressed, leading to local enhancement of turbulence
\citep{RobG12,Lee15,Xug20}.
The role of the adiabatically amplified turbulence in dynamo generation and reconnection diffusion of magnetic fields 
deserves a detailed study. 
\\
\\
\\
We thank Chris McKee for extensive discussions on the
nature of turbulent dynamo and reconnection diffusion as
well as sharing with us the manuscript (M+20)
prior to its publication.
M+20 and discussions with him stimulated our investigation 
of the problem.
S.X. acknowledges the support for Program number HST-HF2-51400.001-A provided by NASA through a grant from the Space Telescope Science Institute, which is operated by the Association of Universities for Research in Astronomy, Incorporated, under NASA contract NAS5-26555.
A.L. acknowledges the support from grants
NSF AST1816234, NASA TCAN 144AAG1967, and NASA  ATP AAH7546.
Flatiron  Institute  is  supported  by  the  Simons Foundation.

\bibliographystyle{apj.bst}
\bibliography{xu}

\begin{thebibliography}{86}
\expandafter\ifx\csname natexlab\endcsname\relax\def\natexlab#1{#1}\fi

\bibitem[{{Batchelor}(1950)}]{Batc50}
{Batchelor}, G.~K. 1950, Royal Society of London Proceedings Series A, 201, 405

\bibitem[{{Beck}(2015)}]{Beck15}
{Beck}, R. 2015, \aapr, 24, 4

\bibitem[{{Beresnyak}(2012)}]{Bere11}
{Beresnyak}, A. 2012, Physical Review Letters, 108, 035002

\bibitem[{{Beresnyak}(2014)}]{Bere14}
---. 2014, \apjl, 784, L20

\bibitem[{{Biermann}(1950)}]{Bierm50}
{Biermann}, L. 1950, Zeitschrift Naturforschung Teil A, 5, 65

\bibitem[{{Brandenburg} \& {Rempel}(2019)}]{Bran19}
{Brandenburg}, A., \& {Rempel}, M. 2019, \apj, 879, 57

\bibitem[{{Brandenburg} \& {Subramanian}(2005)}]{Bran05}
{Brandenburg}, A., \& {Subramanian}, K. 2005, \physrep, 417, 1

\bibitem[{{Cho} \& {Vishniac}(2000{\natexlab{a}})}]{CV00}
{Cho}, J., \& {Vishniac}, E.~T. 2000{\natexlab{a}}, \apj, 539, 273

\bibitem[{{Cho} \& {Vishniac}(2000{\natexlab{b}})}]{Chod00}
---. 2000{\natexlab{b}}, \apj, 538, 217

\bibitem[{{Cho} {et~al.}(2009){Cho}, {Vishniac}, {Beresnyak}, {Lazarian}, \&
  {Ryu}}]{CVB09}
{Cho}, J., {Vishniac}, E.~T., {Beresnyak}, A., {Lazarian}, A., \& {Ryu}, D.
  2009, \apj, 693, 1449

\bibitem[{{Crutcher}(2012)}]{Cru12}
{Crutcher}, R.~M. 2012, \araa, 50, 29

\bibitem[{{Eyink} {et~al.}(2013){Eyink}, {Vishniac}, {Lalescu}, {Aluie},
  {Kanov}, {B{\"u}rger}, {Burns}, {Meneveau}, \& {Szalay}}]{Eyin13}
{Eyink}, G., {et~al.} 2013, \nat, 497, 466

\bibitem[{{Eyink}(2010)}]{Eyi10}
{Eyink}, G.~L. 2010, \pre, 82, 046314

\bibitem[{{Eyink}(2015)}]{Ey15}
---. 2015, \apj, 807, 137

\bibitem[{{Eyink} {et~al.}(2011){Eyink}, {Lazarian}, \& {Vishniac}}]{Eyink2011}
{Eyink}, G.~L., {Lazarian}, A., \& {Vishniac}, E.~T. 2011, \apj, 743, 51

\bibitem[{{Federrath}(2016)}]{Fed16}
{Federrath}, C. 2016, Journal of Plasma Physics, 82, 535820601

\bibitem[{{Federrath} {et~al.}(2011{\natexlab{a}}){Federrath}, {Chabrier},
  {Schober}, {Banerjee}, {Klessen}, \& {Schleicher}}]{Fede11}
{Federrath}, C., {Chabrier}, G., {Schober}, J., {Banerjee}, R., {Klessen},
  R.~S., \& {Schleicher}, D.~R.~G. 2011{\natexlab{a}}, Physical Review Letters,
  107, 114504

\bibitem[{{Federrath} {et~al.}(2014){Federrath}, {Schober}, {Bovino}, \&
  {Schleicher}}]{Fed14}
{Federrath}, C., {Schober}, J., {Bovino}, S., \& {Schleicher}, D. R.~G. 2014,
  \apjl, 797, L19

\bibitem[{{Federrath} {et~al.}(2011{\natexlab{b}}){Federrath}, {Sur},
  {Schleicher}, {Banerjee}, \& {Klessen}}]{Feder11}
{Federrath}, C., {Sur}, S., {Schleicher}, D.~R.~G., {Banerjee}, R., \&
  {Klessen}, R.~S. 2011{\natexlab{b}}, \apj, 731, 62

\bibitem[{{Girichidis} {et~al.}(2014){Girichidis}, {Konstandin}, {Whitworth},
  \& {Klessen}}]{Giri14}
{Girichidis}, P., {Konstandin}, L., {Whitworth}, A.~P., \& {Klessen}, R.~S.
  2014, \apj, 781, 91

\bibitem[{{Goldreich} \& {Sridhar}(1995)}]{GS95}
{Goldreich}, P., \& {Sridhar}, S. 1995, \apj, 438, 763

\bibitem[{{Han}(2017)}]{Han17}
{Han}, J.~L. 2017, \araa, 55, 111

\bibitem[{{Haugen} {et~al.}(2004){Haugen}, {Brandenburg}, \& {Dobler}}]{Hau04}
{Haugen}, N.~E., {Brandenburg}, A., \& {Dobler}, W. 2004, \pre, 70, 016308

\bibitem[{{Haugen} {et~al.}(2003){Haugen}, {Brandenburg}, \& {Dobler}}]{Hau03}
{Haugen}, N. E.~L., {Brandenburg}, A., \& {Dobler}, W. 2003, \apjl, 597, L141

\bibitem[{{Hennebelle} \& {Inutsuka}(2019)}]{Hen19}
{Hennebelle}, P., \& {Inutsuka}, S.-i. 2019, Frontiers in Astronomy and Space
  Sciences, 6, 5

\bibitem[{{Jafari} {et~al.}(2018){Jafari}, {Vishniac}, {Kowal}, \&
  {Lazarian}}]{Jaf18}
{Jafari}, A., {Vishniac}, E.~T., {Kowal}, G., \& {Lazarian}, A. 2018, \apj,
  860, 52

\bibitem[{{Kazantsev}(1968)}]{Kaza68}
{Kazantsev}, A.~P. 1968, Soviet Journal of Experimental and Theoretical
  Physics, 26, 1031

\bibitem[{{Klessen}(2019)}]{Kles19}
{Klessen}, R. 2019, {Formation of the first stars}, ed. M.~{Latif} \&
  D.~{Schleicher}, 67--97

\bibitem[{{Kowal} {et~al.}(2017){Kowal}, {Falceta-Gon{\c c}alves}, {Lazarian},
  \& {Vishniac}}]{Kow17}
{Kowal}, G., {Falceta-Gon{\c c}alves}, D.~A., {Lazarian}, A., \& {Vishniac},
  E.~T. 2017, \apj, 838, 91

\bibitem[{{Kowal} {et~al.}(2020){Kowal}, {Falceta-Gon{\c{c}}alves}, {Lazarian},
  \& {Vishniac}}]{Kow20}
{Kowal}, G., {Falceta-Gon{\c{c}}alves}, D.~A., {Lazarian}, A., \& {Vishniac},
  E.~T. 2020, \apj, 892, 50

\bibitem[{{Kraichnan} \& {Nagarajan}(1967)}]{Krai67}
{Kraichnan}, R.~H., \& {Nagarajan}, S. 1967, Physics of Fluids, 10, 859

\bibitem[{{Kronberg}(2016)}]{Kron16}
{Kronberg}, P.~P. 2016, {Cosmic Magnetic Fields}

\bibitem[{{Krumholz} \& {Federrath}(2019)}]{Krum19}
{Krumholz}, M.~R., \& {Federrath}, C. 2019, Frontiers in Astronomy and Space
  Sciences, 6, 7

\bibitem[{{Kulsrud} \& {Anderson}(1992)}]{KulA92}
{Kulsrud}, R.~M., \& {Anderson}, S.~W. 1992, \apj, 396, 606

\bibitem[{{Lalescu} {et~al.}(2015){Lalescu}, {Shi}, {Eyink}, {Drivas},
  {Vishniac}, \& {Lazarian}}]{Lal15}
{Lalescu}, C.~C., {Shi}, Y.-K., {Eyink}, G.~L., {Drivas}, T.~D., {Vishniac},
  E.~T., \& {Lazarian}, A.~e. 2015, \prl, 115, 025001

\bibitem[{{Latif} {et~al.}(2014){Latif}, {Schleicher}, \& {Schmidt}}]{Lat14}
{Latif}, M.~A., {Schleicher}, D.~R.~G., \& {Schmidt}, W. 2014, \mnras, 440,
  1551

\bibitem[{{Latif} {et~al.}(2013){Latif}, {Schleicher}, {Schmidt}, \&
  {Niemeyer}}]{Lat13}
{Latif}, M.~A., {Schleicher}, D.~R.~G., {Schmidt}, W., \& {Niemeyer}, J. 2013,
  \mnras, 432, 668

\bibitem[{{Lazarian}(2005)}]{Laz05}
{Lazarian}, A. 2005, in American Institute of Physics Conference Series, Vol.
  784, Magnetic Fields in the Universe: From Laboratory and Stars to Primordial
  Structures., ed. E.~M. {de Gouveia dal Pino}, G.~{Lugones}, \& A.~{Lazarian},
  42--53

\bibitem[{{Lazarian}(2014)}]{Laz14r}
{Lazarian}, A. 2014, \ssr, 181, 1

\bibitem[{{Lazarian} {et~al.}(2012){Lazarian}, {Esquivel}, \&
  {Crutcher}}]{LEC12}
{Lazarian}, A., {Esquivel}, A., \& {Crutcher}, R. 2012, \apj, 757, 154

\bibitem[{{Lazarian} {et~al.}(2020){Lazarian}, {Eyink}, {Jafari}, {Kowal},
  {Li}, {Xu}, \& {Vishniac}}]{LaR20}
{Lazarian}, A., {Eyink}, G.~L., {Jafari}, A., {Kowal}, G., {Li}, H., {Xu}, S.,
  \& {Vishniac}, E.~T. 2020, Physics of Plasmas, 27, 012305

\bibitem[{{Lazarian} \& {Vishniac}(1999)}]{LV99}
{Lazarian}, A., \& {Vishniac}, E.~T. 1999, \apj, 517, 700

\bibitem[{{Lee} {et~al.}(2015){Lee}, {Chang}, \& {Murray}}]{Lee15}
{Lee}, E.~J., {Chang}, P., \& {Murray}, N. 2015, \apj, 800, 49

\bibitem[{{Machida} \& {Doi}(2013)}]{Mac13}
{Machida}, M.~N., \& {Doi}, K. 2013, \mnras, 435, 3283

\bibitem[{{Marinacci} {et~al.}(2018){Marinacci}, {Vogelsberger}, {Pakmor},
  {Torrey}, {Springel}, {Hernquist}, {Nelson}, {Weinberger}, {Pillepich},
  {Naiman}, \& {Genel}}]{Mar18}
{Marinacci}, F., {et~al.} 2018, \mnras, 480, 5113

\bibitem[{{Maron} \& {Cowley}(2001)}]{Mar01}
{Maron}, J., \& {Cowley}, S. 2001, astro-ph/0111008, astro

\bibitem[{{Maron} \& {Goldreich}(2001)}]{MG01}
{Maron}, J., \& {Goldreich}, P. 2001, \apj, 554, 1175

\bibitem[{{McKee} \& {Ostriker}(2007)}]{Mckee_Ostriker2007}
{McKee}, C.~F., \& {Ostriker}, E.~C. 2007, \araa, 45, 565

\bibitem[{{McKee} {et~al.}(2020){McKee}, {Stacy}, \& {Li}}]{Mc20}
{McKee}, C.~F., {Stacy}, A., \& {Li}, P.~S. 2020, arXiv:2006.14607

\bibitem[{{Nixon} \& {Pringle}(2019)}]{Nix19}
{Nixon}, C.~J., \& {Pringle}, J.~E. 2019, \na, 67, 89

\bibitem[{{Parker}(1957)}]{Park57}
{Parker}, E.~N. 1957, \jgr, 62, 509

\bibitem[{{Parker}(1963)}]{Par63}
---. 1963, \apjs, 8, 177

\bibitem[{{Parker}(1979)}]{Park79}
---. 1979, {Cosmical magnetic fields. Their origin and their activity}

\bibitem[{{Petschek}(1964)}]{Pet64}
{Petschek}, H.~E. 1964, {Magnetic Field Annihilation}, Vol.~50, 425

\bibitem[{{Pudritz} \& {Ray}(2019)}]{Pud19}
{Pudritz}, R.~E., \& {Ray}, T.~P. 2019, Frontiers in Astronomy and Space
  Sciences, 6, 54

\bibitem[{{Quashnock} {et~al.}(1989){Quashnock}, {Loeb}, \& {Spergel}}]{Qua89}
{Quashnock}, J.~M., {Loeb}, A., \& {Spergel}, D.~N. 1989, \apjl, 344, L49

\bibitem[{{Robertson} \& {Goldreich}(2012)}]{RobG12}
{Robertson}, B., \& {Goldreich}, P. 2012, \apjl, 750, L31

\bibitem[{{Ruzmaikin} {et~al.}(1989){Ruzmaikin}, {Sokolov}, \&
  {Shukurov}}]{Ruz89}
{Ruzmaikin}, A., {Sokolov}, D., \& {Shukurov}, A. 1989, \mnras, 241, 1

\bibitem[{{Ryu} {et~al.}(2008){Ryu}, {Kang}, {Cho}, \& {Das}}]{Ryu08}
{Ryu}, D., {Kang}, H., {Cho}, J., \& {Das}, S. 2008, Science, 320, 909

\bibitem[{{Santos-Lima} {et~al.}(2020){Santos-Lima}, {Guerrero}, {de Gouveia
  Dal Pino}, \& {Lazarian}}]{San20}
{Santos-Lima}, R., {Guerrero}, G., {de Gouveia Dal Pino}, E.~M., \& {Lazarian},
  A. 2020, arXiv:2005.07775, arXiv:2005.07775

\bibitem[{{Santos-Lima} {et~al.}(2010){Santos-Lima}, {Lazarian}, {de Gouveia
  Dal Pino}, \& {Cho}}]{Sant10}
{Santos-Lima}, R., {Lazarian}, A., {de Gouveia Dal Pino}, E.~M., \& {Cho}, J.
  2010, \apj, 714, 442

\bibitem[{{Schekochihin} {et~al.}(2002{\natexlab{a}}){Schekochihin}, {Cowley},
  {Maron}, \& {Malyshkin}}]{Schek02}
{Schekochihin}, A., {Cowley}, S., {Maron}, J., \& {Malyshkin}, L.
  2002{\natexlab{a}}, \pre, 65, 016305

\bibitem[{{Schekochihin} \& {Cowley}(2007)}]{SchC07}
{Schekochihin}, A.~A., \& {Cowley}, S.~C. 2007, {Turbulence and Magnetic Fields
  in Astrophysical Plasmas}, ed. S.~{Molokov}, R.~{Moreau}, \& H.~K. {Moffatt}
  (Springer), 85

\bibitem[{{Schekochihin} {et~al.}(2002{\natexlab{b}}){Schekochihin}, {Cowley},
  {Hammett}, {Maron}, \& {McWilliams}}]{Sch02}
{Schekochihin}, A.~A., {Cowley}, S.~C., {Hammett}, G.~W., {Maron}, J.~L., \&
  {McWilliams}, J.~C. 2002{\natexlab{b}}, New Journal of Physics, 4, 84

\bibitem[{{Schekochihin} {et~al.}(2004){Schekochihin}, {Cowley}, {Taylor},
  {Maron}, \& {McWilliams}}]{Sch04}
{Schekochihin}, A.~A., {Cowley}, S.~C., {Taylor}, S.~F., {Maron}, J.~L., \&
  {McWilliams}, J.~C. 2004, \apj, 612, 276

\bibitem[{{Schlickeiser} {et~al.}(2018){Schlickeiser}, {Kolberg}, \&
  {Yoon}}]{Sch18}
{Schlickeiser}, R., {Kolberg}, U., \& {Yoon}, P.~H. 2018, \apj, 857, 29

\bibitem[{{Schl{\"u}ter} \& {Biermann}(1950)}]{Sch50}
{Schl{\"u}ter}, A., \& {Biermann}, L. 1950, Zeitschrift Naturforschung Teil A,
  5, 237

\bibitem[{{Schober} {et~al.}(2012{\natexlab{a}}){Schober}, {Schleicher},
  {Federrath}, {Glover}, {Klessen}, \& {Banerjee}}]{SchoSch12}
{Schober}, J., {Schleicher}, D., {Federrath}, C., {Glover}, S., {Klessen},
  R.~S., \& {Banerjee}, R. 2012{\natexlab{a}}, \apj, 754, 99

\bibitem[{{Schober} {et~al.}(2012{\natexlab{b}}){Schober}, {Schleicher},
  {Federrath}, {Klessen}, \& {Banerjee}}]{Scho12}
{Schober}, J., {Schleicher}, D., {Federrath}, C., {Klessen}, R., \& {Banerjee},
  R. 2012{\natexlab{b}}, \pre, 85, 026303

\bibitem[{{Schober} {et~al.}(2015){Schober}, {Schleicher}, {Federrath},
  {Bovino}, \& {Klessen}}]{Scho15}
{Schober}, J., {Schleicher}, D.~R.~G., {Federrath}, C., {Bovino}, S., \&
  {Klessen}, R.~S. 2015, \pre, 92, 023010

\bibitem[{{Sharda} {et~al.}(2020){Sharda}, {Federrath}, \& {Krumholz}}]{Shar20}
{Sharda}, P., {Federrath}, C., \& {Krumholz}, M.~R. 2020, arXiv:2002.11502,
  arXiv:2002.11502

\bibitem[{{Sigl} {et~al.}(1997){Sigl}, {Olinto}, \& {Jedamzik}}]{Sigl97}
{Sigl}, G., {Olinto}, A.~V., \& {Jedamzik}, K. 1997, \prd, 55, 4582

\bibitem[{{Spitzer}(1968)}]{Spit68}
{Spitzer}, L. 1968, {Diffuse matter in space}

\bibitem[{{Subramanian}(1998)}]{Subra98}
{Subramanian}, K. 1998, \mnras, 294, 718

\bibitem[{{Subramanian}(1999)}]{Sub99}
---. 1999, \prl, 83, 2957

\bibitem[{{Subramanian}(2003)}]{Sub03}
---. 2003, \prl, 90, 245003

\bibitem[{{Sur} {et~al.}(2012){Sur}, {Federrath}, {Schleicher}, {Banerjee}, \&
  {Klessen}}]{Sur12}
{Sur}, S., {Federrath}, C., {Schleicher}, D.~R.~G., {Banerjee}, R., \&
  {Klessen}, R.~S. 2012, \mnras, 423, 3148

\bibitem[{{Sur} {et~al.}(2010){Sur}, {Schleicher}, {Banerjee}, {Federrath}, \&
  {Klessen}}]{Sur10}
{Sur}, S., {Schleicher}, D.~R.~G., {Banerjee}, R., {Federrath}, C., \&
  {Klessen}, R.~S. 2010, \apjl, 721, L134

\bibitem[{{Sweet}(1958)}]{Swe58}
{Sweet}, P.~A. 1958, in IAU Symposium, Vol.~6, Electromagnetic Phenomena in
  Cosmical Physics, ed. B.~{Lehnert}, 123

\bibitem[{{Turk} {et~al.}(2012){Turk}, {Oishi}, {Abel}, \& {Bryan}}]{Tur12}
{Turk}, M.~J., {Oishi}, J.~S., {Abel}, T., \& {Bryan}, G.~L. 2012, \apj, 745,
  154

\bibitem[{{Vogel}(2016)}]{Vog16}
{Vogel}, M. 2016, Contemporary Physics, 57, 134

\bibitem[{{Vogt} \& {En{\ss}lin}(2005)}]{Vog05}
{Vogt}, C., \& {En{\ss}lin}, T.~A. 2005, \aap, 434, 67

\bibitem[{{Xu} {et~al.}(2019){Xu}, {Garain}, {Balsara}, \& {Lazarian}}]{Xud19}
{Xu}, S., {Garain}, S.~K., {Balsara}, D.~S., \& {Lazarian}, A. 2019, \apj, 872,
  62

\bibitem[{{Xu} \& {Lazarian}(2016)}]{XL16}
{Xu}, S., \& {Lazarian}, A. 2016, \apj, 833, 215

\bibitem[{{Xu} \& {Lazarian}(2017)}]{XuL17}
---. 2017, \apj, 850, 126

\bibitem[{{Xu} \& {Lazarian}(2020)}]{Xug20}
---. 2020, \apj, 890, 157

\end{thebibliography}

\end{document}